\documentclass[aps,prl,twocolumn,groupedaddress,showpacs]{revtex4}
\usepackage{graphicx}

\begin{document}

\title{Statistics of unstable periodic orbits of a chaotic dynamical
system with a large number of degrees of freedom}

\author{Mitsuhiro Kawasaki}
\email[mkawasaki@eng.niigata-u.ac.jp]{}
\affiliation{Department of Materials Science and Technology, Niigata
University, Niigata 950-2181, Japan}
\author{Shin-ichi Sasa}
\email[sasa@jiro.c.u-tokyo.ac.jp]{}
\affiliation{Department of Pure and Applied Sciences, University of
Tokyo, Tokyo 153-8902, Japan}

\date{\today}

\begin{abstract}
For a simple model of chaotic dynamical systems with a large number of 
degrees of  freedom, we find that there is an ensemble 
of unstable periodic orbits (UPOs) with the special property 
that the expectation values 
of macroscopic quantities can be calculated using only one UPO 
sampled from the ensemble. Evidence to support this conclusion is 
obtained by generating the ensemble by Monte Carlo calculation 
for a statistical mechanical model described 
by a space-time  Hamiltonian
that is expressed in terms of Floquet exponents of UPOs. This result 
allows us to interpret the recent interesting discovery 
that statistical properties of turbulence can be obtained from only one
UPO [G. Kawahara and S. Kida, J.\ Fluid\ Mech. {\bf 449}, 291 (2001); 
S. Kato and M. Yamada, Phys.\ Rev.\ E {\bf 68}, 025302(R)(2003)].
\end{abstract}


\maketitle

A wide range of systems, including fluid turbulence and ecosystems, 
can be described by chaotic dynamical systems (CDSs) with a large 
number of degrees of freedom. Because the evolution equations describing 
them are nonlinear and possess many degrees of freedom, 
analyses employing various theoretical tools fail. 
For instance, it is difficult to characterize 
quantitatively the intermittency of fluid turbulence 
using perturbative expansion methods applied to the Navier-Stokes equation
\cite{mccomb1990}.

In contrast, for CDSs consisting of assemblies of molecules at 
equilibrium, equilibrium statistical mechanics 
provides a powerful framework to predict macroscopic properties 
without the need to analyze a Hamiltonian equation for a large number of
molecules. The success of equilibrium 
statistical mechanics relies on its probabilistic description: In order
to predict macroscopic properties, the exact probability distribution
for states is not needed. Rather, the existence of many degrees of
freedom allows a tractable distribution to reproduce 
macroscopic properties correctly. 
When considering macroscopic properties of a CDS  
with many degrees of freedom, it is tempting to think that 
a probabilistic approach other than 
analyzing evolution equations can be employed by finding a useful probability  
measure for the system. 

Although it would be extremely difficult to find  a useful measure 
for general CDSs, it was suggested recently that an ensemble of states 
that describes macroscopic properties of CDSs can be
constructed from a special unstable periodic orbit (UPO)
\cite{kawahara2001,kato2003}. 
The first study of such a UPO demonstrated that the spatial profiles 
of the mean and variance of the velocity in minimal wall turbulence can  
be extracted from only one  UPO of the Navier-Stokes equation 
\cite{kawahara2001}. Subsequently, the scaling exponents 
of  velocity fluctuations, which characterize the intermittency 
of a turbulent velocity field, was found to be obtained from 
only one UPO in the GOY shell model \cite{kato2003}.

To obtain both of these interesting results, 
the UPOs were found in numerical computations 
such as the Newton-Raphson method applied to a function of 15,422 variables 
\cite{kawahara2001} and following the destabilization of a limit cycle
resulting from a bifurcation process in the GOY model with 24 degrees of 
freedom \cite{kato2003}. Because there are other UPOs which yield 
properties that differ from those of the special UPOs, 
it can be considered that these special UPOs were selected 
out of infinitely many UPOs under some criterion. However, such a criterion 
for choosing UPOs is still unknown, because in the previous works the 
initial value for the Newton-Raphson method and the path of the 
destabilization of the limit cycle were found by trial and error. 
Given this situation, the objective of the present work is to formulate 
a method to construct an ensemble of 
special UPOs such that macroscopic properties 
can be obtained with high accuracy even when only one UPO element of 
the ensemble is used.

In this Letter, we present analysis in which such a UPO ensemble is
obtained for a CDS with many degrees of freedom. 
Note that here, the term ``macroscopic quantities''
refers to quantities obtained by taking the average over many degrees of 
freedom of the system. Furthermore, the term ``macroscopic properties'' 
is used to mean 
the leading terms of the expectation values of the macroscopic
quantities with respect to the natural invariant measure 
when the number of degrees of freedom is large.

\paragraph{Model:}

Here, we explain the model analyzed in the present work. 
First, we note that it is not necessary to study turbulence for our 
present purpose. Rather, it may be preferable to analyze a simple model 
for which many UPOs can be found easily. 
With this idea, we study a coupled map lattice (CML) 
proposed by Sakaguchi \cite{sakaguchi1988}, which we now describe. 
Let $(x_i,\Delta_i) \in 
[-1,1]\times [-1,1]$  be dynamical variables defined 
on the $i$-th site of a one-dimensional lattice consisting of $N$
lattice points numbered from $0$ to $N-1$ \cite{note1}. 
For later convenience, the variable ${s^t}_i$, which is
called the ``spin'', is defined as
\begin{eqnarray}
{s}_i \equiv \left\{ 
	    \begin{array}{ll}
	     +1 & (-1 \leq {x}_i < {{\triangle}}_i), \\
	     -1 & ({{\triangle}}_i \leq {x}_i \leq 1 ).
	    \end{array}
	   \right.
 \label{eq:spin}
\end{eqnarray}
The time evolution of $(x_i,\Delta_i) $ is given by 
\begin{equation}
({x^{t+1}}_i, {\triangle^{t+1}}_i) = 
(f({x^t}_i, {\triangle^t}_i), 
\tanh\left[\frac{k}{2}({s^t}_{i-1}+{s^t}_{i+1})\right]) 
\end{equation}
for $i=t+1$  (mod $2$) and $({x^{t+1}}_i, {\triangle^{t+1}}_i) = 
({x^t}_i, {\triangle^t}_i)$ for $i=t$  (mod $2$). 
Here $k$ is a positive parameter, and the local map
$f({x^t}_i, {\triangle^t}_i)$ is the Bernoulli map, given by 
\begin{equation}
 f({x^t}_i, {\triangle^t}_i) \equiv 
\frac{2 ({x^t}_i+{s^t}_i)}{1+{s^t}_i{{\triangle}^t}_i}
-{s^t}_i .
\label{eq:local-map}
\end{equation}
Note that the discontinuous point of the map $f(x, \triangle)$ is 
located at $x=\triangle$, and the spin variable, $s_i$, is identified
to a symbol of the symbolic dynamical system for the local map 
$f(x_i, \triangle_i)$. Hereafter, we call the above CML the ``Bernoulli 
CML'' and denote it $X^{t+1}=F(X^t)$, where $X^t\equiv \{{x^t}_i, 
{\triangle^t}_i\}^{N-1}_{i=0}$. 

The Bernoulli CML has interesting features, 
as demonstrated by Sakaguchi \cite{sakaguchi1988}. 
First, the natural 
invariant measure for spin configurations coincides with the canonical
distribution for an Ising spin Hamiltonian. In order to express this 
fact explicitly, we define $J(s)$  as  the set of states 
$\{x_i, \triangle_i\}_{i=0}^{N-1}$ corresponding to the spin 
configuration $s\equiv \{s_i\}^{N-1}_{i=0}$. 
Then, it is found that the natural invariant 
measure on the set $J(s)$ can be written as 
$
\mu(J(s)) = {\rm const.} \times 
\exp[(k/2) \sum_{i=0}^{N-1} s_i s_{i+1}].
$
Furthermore, when the initial probability distribution of states is the 
natural invariant measure, it is known that the transition probability 
$T(s|s')$  from one spin configuration $s$ to another $s'$ is
given by 
$
 T(s^{2t}|s^{2t+1}) = \prod_{i=0}^{N/2-1} (1+{s^{2t+1}}_{2i+1}
  {\triangle^{2t+1}}_{2i+1})/2, 
 T(s^{2t+1}|s^{2t+2}) = \prod_{i=0}^{N/2-1} (1+{s^{2t+2}}_{2i}
  {\triangle^{2t+2}}_{2i})/2.
$


\paragraph{A UPO ensemble:}

We now proceed to construct a UPO ensemble for the Bernoulli CML with
the property that the 
expectation values of macroscopic quantities are obtained with high
accuracy 
even when only one UPO element of the ensemble is used. 
As the first step in this construction, we demonstrate that there is 
a one-to-one correspondence between symbol sequences and UPOs. 
Suppose that a symbol sequence $[s]\equiv ({s^0}, {s^1}, \ldots,
{s^{p-1}})$
is given. We then attempt to find a periodic point $X^0 \equiv \{{x^0}_i,
{\triangle^0}_i\}_{i=0}^{N-1}$ corresponding to $[s]$. 
The $N$-tuple $\{{\triangle^0}_i \}^{N-1}_{i=0}$ can be 
determined directly from the given symbol sequence by definition. 
A component of $X^p \equiv F^p (X^0)$ is given by 
$
 {x^p}_i 
 = \prod_{t=0}^{p/2-1}{a^{2t}}_i
  {x^0}_i+\sum_{t'=1}^{p/2-1}\prod_{t=t'}^{p/2-1}{a^{2t}}_i
{s^{2t'-2}}_i({a^{2t'-2}}_i-1)+{s^{p-2}}_i({a^{p-2}}_i-1),
$
where it is assumed that $p$ is an even number and 
${a^t}_i\equiv 2/(1+{s^t}_i {\triangle^t}_i)$. Because 
$\prod_{t=0}^{p/2-1}{a^t}_i\neq 1$  for an arbitrary symbol sequence,
we can uniquely determine the point satisfying the periodicity condition
${x^0}_i={x^p}_i$. In this way,
we can find a point $X^0 \equiv \{{x^0}_i, {\triangle^0}_i \}_{i=0}^{N-1}$ 
such that $X^0=X^p$ for a given symbol sequence $[s]$. Conversely,
it should be confirmed that the symbol sequence generated from
the point $X^0$ coincides with $[s]$. We numerically confirmed
this for $10^6$ symbol sequences generated randomly. 
On the basis of these results, we conclude that 
there exists  a single UPO corresponding to an 
arbitrary symbol sequence.  

Because of the one-to-one correspondence between UPOs and symbol sequences,
we can obtain a UPO ensemble through the construction of an ensemble of symbol
sequences. One natural possibility 
for a probability measure (PM) on symbol sequences is the frequency
distribution of symbol sequences in the case that 
the initial probability distribution of 
states is given by the natural invariant measure. The
frequency distribution of a symbol sequence $[s]$ is 
expressed as $P([s])=\mu(J(s^0)) T(s^0|s^1)T(s^1|s^2)\cdots
T(s^{p-1}|s^p)$. 
Although this expression appears to take a simple form, 
it is not easy to derive the natural invariant measure for most
dynamical systems. In order to analyze a wide variety of dynamical
systems, it is convenient to use a more tractable PM that yields the 
same macroscopic properties as those obtained from $P([s])$.
As one such possibility, we consider the PM $Q([s])$ obtained by
replacing the natural invariant measure $\mu$ in $P([s])$ with a
constant $Z^{-1}$. This PM can be rewritten as 
\begin{equation}
 Q([s]) = \exp[-\ln |D^{(u)} F^p(X^0)|]/Z.
 \label{eq:sampling}
\end{equation}
Here, $|D^{(u)}F^p(X^0)|$ 
is the absolute value of the product of the eigenvalues in the expanding
directions of the Jacobi matrix for the map given by $p$ iterations 
of the coupled map $F$ evaluated at the periodic point $X^0$ 
corresponding to the symbol sequence $[s]$, and $Z$ is a normalization 
constant. The expression of frequency distribution of a symbol sequence
has been obtained in Ref. \cite{grassberger1988}. 
The PM defined by Eq.\  (\ref{eq:sampling}) can be calculated more
easily for any CDS than that defined by $P([s])$.  

\paragraph{Numerical demonstration:}

Now we describe the numerical demonstration that a macroscopic 
property for the CML can be determined with high accuracy even 
when only one UPO element of the ensemble determined by 
Eq.\ (\ref{eq:sampling}) is used to provide ensembles 
of states.  First, we note that the sampling of UPOs according to 
Eq.\ (\ref{eq:sampling}) can be carried out using the Monte Carlo method. 
In this procedure, a symbol sequence $[s]$ 
is regarded as a spin configuration on a $1+1$-dimensional
$p \times N$ lattice, where $p$ and $N$ correspond to the time and space
directions, respectively. In addition, the PM given by Eq.\
(\ref{eq:sampling})
is of the same form as the canonical distribution. Hence, 
the UPO ensemble we seek can be 
generated with the Metropolis algorithm, regarding 
$\ln |D^{(u)} F^p(X^0)|$ as the ``Hamiltonian'' of a $1+1$-dimensional 
Ising model with spins $[s]$. 
In order to obtain UPOs, periodic boundary conditions are imposed at 
the boundaries of the lattice in the time direction, $t = 0$ and $t = p-1$.

\begin{figure}
\includegraphics[width=8cm,keepaspectratio]{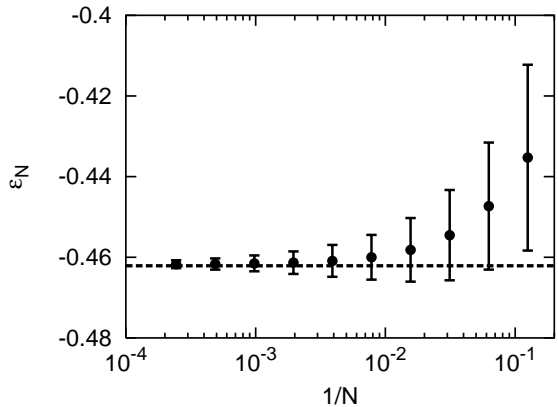}
\caption{\label{fig:convergence}
Averages (solid circles) and  standard deviations (error bars) of 
$\varepsilon_N$ evaluated from 50,000 samples of UPOs 
for the case $N=2^n \ (3 \le n \le 12)$, $k=1, p=64$. 
The broken line represents the exact value in the limit
 $N\rightarrow\infty$, $-\tanh(1/2)$.
}
\end{figure}

For each UPO of period $p$, an ensemble of spin configurations $s$, 
which can be regarded as an ensemble of states, can be constructed by 
assuming the equal weight probability $1/p$ on each spin configuration
of the sequence $[s]$ corresponding to the UPO. Using this ensemble
associated with one UPO, we consider the ensemble average of the 
macroscopic quantity $-\sum_{i=0}^{N-1}{s}_i{s}_{i+1}/N$, given by 
$
\varepsilon_N 
\equiv -\sum_{t=0}^{p-1}\sum_{i=0}^{N-1}{s^t}_i{s^t}_{i+1}/(N p).
$
In Fig.\ \ref{fig:convergence},
the average and standard deviation of $\varepsilon_N$ calculated 
from 50,000 samples of UPOs generated by the Monte Carlo method 
are displayed  for several values of $N$. 
It is seen that as $N$ increases, the standard deviation 
approaches zero and the average value of $\varepsilon_N$ converges 
to $-\tanh(k/2)$, which is the value calculated from the natural invariant
measure $\mu(J(s))$. 
This implies that in the limit
$N\rightarrow\infty$, the expectation value of 
$-\sum_{i=0}^{N-1}{s}_i{s}_{i+1}/N$ with respect to the natural
invariant measure can be calculated from the ensemble of states 
obtained from only one UPO, because the
discrepancy between the value calculated from one UPO 
and that from another UPO has turned out to be ignored. 
This is the main result of the present Letter. 

\paragraph{Single-UPO description:}

It is worthwhile to note that our result can be understood for a wide
class of CDSs with large degrees of freedom. First, let us consider
relaxation processes to the steady state in such a CDS provided that a
smooth initial measure is given. 
Then, the expectation value of a quantity that is a function of a
symbol, $A(s)$, converges to the same value in a relaxation time
$\tau_A$, irrespective of initial measures. Thus, the asymptotic value
for the case that the initial measure is assumed to be flat, where the
frequency distribution of the symbol sequence is equal to $Q([s])$,
coincides with the value for the case that the initial measure is given
by the natural measure $\mu$, whose value is nothing but the expectation
value of $A(s)$ with respect to $\mu$. 
This is expressed simply as 
$\langle A(s^t) \rangle_{Q}
\equiv \sum_{[s]}Q([s])A(s^t) \simeq \langle A(s) \rangle_\mu\equiv
\sum_s \mu(J(s)) A(s)$ in the time range $t>\tau_A$. 

Next, by using this generic property, 
we can estimate the expectation value of the time average of $A(s)$ as 
$
\left\langle \sum_{t=0}^{p-1}A(s^t)/p \right\rangle_Q
\simeq \sum_{t=0}^{\tau_A-1}\langle A(s^t)
\rangle_Q/p+(p-\tau_A)\langle A(s)
\rangle_{\mu}/p.
$
When the period $p$ is taken to be long compared to the 
relaxation time $\tau_A$, we obtain $\left\langle
\sum_{t=0}^{p-1}A(s^t)/p\right\rangle_Q\simeq \langle A(s)
\rangle_\mu$. 

Finally, we consider the case where $A(s)$ is a macroscopic quantity such as
$\varepsilon_N$. The law of large numbers asserts that fluctuations
of such a quantity become negligible when the degrees of freedom is
large and steady state is settled. Hence, for almost all symbol
sequence $[s]$ sampled according to $Q([s])$, $\sum_{t=0}^{p-1}A(s^t)/p
\simeq \left\langle\sum_{t=0}^{p-1}A(s^t)/p
\right\rangle_{Q}$. 

Consequently, 
$\sum_{t=0}^{p-1}A(s^t)/p \simeq \langle A(s)
\rangle_\mu$ for almost all $[s]$ with respect to $Q([s])$. 
It means that the time average along single UPO sampled according
to $Q([s])$ provides a good approximate value of 
$\langle A_N(s) \rangle_\mu$. 
Since the discussion presented above is based on generic properties 
of CDSs with many degrees of freedom, it is highly
expected that description of macroscopic properties with single UPO can
be applied for a wide class of systems.

\paragraph{Interpretation of the UPO description of turbulence:}

We now interpret the UPO description of turbulence
\cite{kawahara2001,kato2003} based on our result. 
In particular, we discuss the relationship between the 
UPOs employed in Refs.\ \cite{kawahara2001, kato2003} 
and the UPOs sampled according to Eq.\ (\ref{eq:sampling}). 

In the methods of searching for UPOs employed in the two previous works, 
a trial initial state point is first chosen near a true periodic point $X^0$ 
in some way, and the time development of the state
is traced by numerical integration of the governing equation. 
Then, the initial point is recognized as a 
periodic point if the distance between the orbit and 
the trial initial state point at the time that the orbit returns 
closest to the trial initial state point is less than some 
threshold value $\delta$. 

For simplicity, suppose that a set of trial initial state points 
are scattered within a hypersphere of radius $\delta_0$ centered at a 
periodic point $X^0$ of period $p$. 
Then, this set expands in the unstable direction as the system evolves
in time. Specifically, assuming that $d_u$ is the dimension of the unstable
manifold and $\delta_0$ is sufficiently small, the volume of the set 
in the unstable direction becomes $\delta^{d_u}_0 |D^{(u)}F^p(X^0)|$ 
after one period of motion along the UPO starting from $X^0$. 
Hence, the probability that the distance between the orbit and the trial
initial state point after one period of motion 
remains less than the threshold $\delta$ 
is estimated as $\delta^{d_u}/(\delta^{d_u}_0 |D^{(u)}F^p(X^0)|)$. 
This implies that the probability of finding numerically 
an approximate periodic point around  $X^0$  is proportional to 
$|D^{(u)}F^p(X^0)|^{-1}$, which is equal to the quantity appearing
in Eq.\ (\ref{eq:sampling}). 
From these considerations, we conjecture that the UPOs found in 
the previous works \cite{kawahara2001, kato2003} were sampled from a
probability measure that takes a form similar to 
Eq.\ (\ref{eq:sampling}) for turbulence. 

In order to demonstrate the correspondence between the single-UPO 
description of turbulence 
and our result that the expectation value of a quantity averaged 
over many degrees of freedom can be obtained from only one UPO, one task
remains: We must show that the quantities analyzed in  
Refs. \cite{kawahara2001,kato2003} were also averaged over many degrees
of freedom. First, the quantities investigated in the case of wall turbulence 
\cite{kawahara2001} were averaged on a plane parallel to the walls. 
Next, consider 
whether the scaling exponents of the turbulent velocity field analyzed 
in Ref.\ \cite{kato2003} can be expressed in an averaged form over 
many degrees of freedom. By making use of notions on the 
energy cascade and self-similarity in the inertial subrange, the
scaling exponent of order $q$, $\zeta_q$, can be derived as $\zeta_q = q/3-\ln
\langle (\epsilon_{j+1}/\epsilon_j)^{q/3} \rangle$, where $\epsilon_j$
is the energy dissipation rate at the $j$-th shell \cite{frisch1995}. 
Self-similarity makes the statistical average $\langle \cdot \rangle$
replaceable with the average over shells in the
inertial subrange as $\langle (\epsilon_{j+1}/\epsilon_j)^{q/3} \rangle
\simeq (1/N) \sum_{j=0}^{N-1} (\epsilon_{j+1}/\epsilon_j)^{q/3}$. 
Thus, the scaling exponent $\zeta_q$ can be expressed as a quantity 
averaged over identically distributed variables. 

One may wonder whether the numbers of statistically independent
variables in the systems studied in Refs. \cite{kawahara2001,kato2003}
are sufficiently large to ensure that the UPOs sampled according to Eq.\ 
(\ref{eq:sampling}) should actually to yield values close to the 
true expectation value. However, as seen in Fig.\
\ref{fig:convergence}, 
the results for the cases $N=8$ and $16$ indeed provide good
approximations 
of the true expectation value. Therefore, 
in the turbulence problems, we believe that the numbers of statistically 
independent variables are sufficient to provide 
good approximations.

\paragraph{Discussion :}

The statistical analysis of symbol sequences for orbits
in dynamical systems is called the ``thermodynamic formalism'' 
\cite{ruelle1978,beck1993}.  The Monte Carlo calculation  
of Eq.\ (\ref{eq:sampling}) can be regarded as a numerical 
realization of the thermodynamic formalism for the Bernoulli CML.
In similar studies, the explicit construction of the space-time 
Hamiltonian for spatially extended dynamical 
systems was proposed in Ref.\ \cite{just1998}, though 
its construction is  based on a natural invariant measure
that happened to be obtained for specific models. 

Let us briefly discuss the range of applicability of our method.
In hyperbolic dynamical systems, employing the periodic orbit
expansion, 
the natural invariant measure $\mu$ on a set $R$ in phase space can be
expressed as 
$
 \mu(R) = \lim_{p\rightarrow\infty} \sum_{j} |D^{(u)} F^p(X_j)|^{-1},
$
where the sum is taken over all the fixed points $X_j$ of 
$F^p$, i.e., UPOs, in the set $R$ \cite{morita1988, 
grebogi1988, cvitanovic2003}. 
From this expanded form, it turns out that  the
weight of a UPO is proportional to $|D^{(u)} F^p(X_j)|^{-1}$.
Because this factor has the same form as Eq.\ (\ref{eq:sampling}),
our method can be applied to general hyperbolic systems.
In related studies, some CDSs with many degrees of freedom 
have been analyzed using the periodic orbit expansion 
technique \cite{politi1992, christiansen1997, lan2003}.  
In contrast to the expansion employing 
many UPOs,  we have demonstrated that only one UPO sampled according 
to Eq.\ (\ref{eq:sampling}) is sufficient to provide a description of
macroscopic quantities. 


In summary, we have demonstrated that 
the macroscopic properties of the Bernoulli CML can be 
calculated with high accuracy using only one UPO sampled 
from the special UPO ensemble described by Eq.\ (\ref{eq:sampling}).

\begin{acknowledgments}
 We acknowledge valuable discussions with Prof.\ Yamada and Prof.\
 Kawahara. 
 One of authors (M. K.) is grateful for the hospitality he received at 
 University of Tokyo. This work was partly supported by a 
 JSPS Research Fellowship for Young Scientists. 
\end{acknowledgments}


\begin{thebibliography}{99}
\bibitem{note1} Sakaguchi analyzed a two-dimensional case. 
A one-dimensional lattice is sufficient for our purpose.
\end{thebibliography}
\end{document}